# Direct observation of the Dirac nodes lifting in semimetallic perovskite SrIrO$_3$ thin films

Z. T. Liu[1], M. Y. Li[1], Q. F. Li[2,3], J. S. Liu[1], W. Li[1], H. F. Yang[1], Q. Yao[1,4,5], C. C. Fan[1], X. G. Wan[2], Z. Wang[1] and D. W. Shen[1,6,*]

[1]State Key Laboratory of Functional Materials for Informatics,
Shanghai Institute of Microsystem and Information Technology (SIMIT),
Chinese Academy of Sciences, Shanghai 200050, China
[2]National Laboratory of Solid State Microstructures and Department of Physics,
National Center of Microstructures and Quantum Manipulation, Nanjing University,
Nanjing 210093, China
[3]Department of Physics, Nanjing University of Information Science &Technology,
Nanjing 210044, China
[4]State Key Laboratory of Surface Physics, Department of Physics, and Advanced
Materials Laboratory, Fudan University, Shanghai 200433, China
[5]Collaborative Innovation Center of Advanced Microstructures, Fudan University,
Shanghai 200433, China
[6]CAS-Shanghai Science Research Center, Shanghai 201203, China

*dwshen@mail.sim.ac.cn

**Perovskite SrIrO$_3$ has long been proposed as an exotic semimetal induced by the interplay between the spin-orbit coupling and electron correlations. However, its low-lying electronic structure is still lacking. We synthesize high-quality perovskite SrIrO$_3$ (100) films by means of oxide molecular beam epitaxy, and then systemically investigate their low energy electronic structure using *in-situ* angle-resolved photoemission spectroscopy. We find that the hole-like bands around *R* and the electron-like bands around *U(T)* intersect the Fermi level simultaneously, providing the direct evidence of the semimetallic ground state in this compound. Comparing with the density functional theory, we discover that the bandwidth of states near Fermi level is extremely small, and there exists a pronounced mixing between the $J_{eff}$ = 1/2 and $J_{eff}$ = 3/2 states. Moreover, our data reveal that the predicted Dirac degeneracy protected by the mirror-symmetry, which was theoretically suggested to be the key to realize the non-trivial topological properties, is actually lifted in perovskite SrIrO$_3$ thin films. Our findings pose**

**strong constraints on the current theoretical models for the 5$d$ iridates.**

## Introduction

Recently, the strongly spin-orbit coupled 5$d$-electron iridates have aroused a great deal of interests[1–12]. Different from the well-studied 3$d$ transition metal oxides, the much larger spin-orbit interaction (SOI) and relatively weak electron correlations are roughly of the same magnitude in iridates, and consequently their delicate interplay has been suggested to host some exotic quantum ground states, e.g., spin-orbit coupled Mott insulators[1,13-15], topological semimetals[4-7], axion insulators[8], correlated topological insulators[4,9–12,16,17], unconventional superconductors[18,19] and spin liquids[20-23].

Among all iridates, the Ruddlesden-Popper (RP) series $Sr_{n+1}Ir_nO_{3n+1}$ have received the special attention. These compounds display distinct properties depending on the different $n$ value. $Sr_2IrO_4$ ($n = 1$) has been regarded as a prototype of the so-called spin-orbit coupled Mott insulator, in which the strong SOI lifts the orbital degeneracy of the $t_{2g}$ bands and results in such a narrow half-filled $J_{eff} = 1/2$ band that even moderate electron correlations could induce a Mott metal-insulator transition (MIT) in this 5$d$-electron system[1,2,13]. As for the $n = 2$ case, $Sr_3Ir_2O_7$ has been proven to form a similar effective $J_{eff} = 1/2$ band but just on the verge of the Mott insulator regime with a smaller band gap[24,25]. Naturally, the $n = \infty$ end-member $SrIrO_3$ was supposed to be on the itinerant side of this family[2,26]. In this regard, this family of compounds could provide a unique opportunity to investigate the MIT in the strong SOI limit. Since $Sr_2IrO_4$ and $Sr_3Ir_2O_7$ have been comprehensively studied, it is crucial to understand the low energy electronic structure of $SrIrO_3$, so that a complete picture of this MIT could be obtained. Besides, theories based on tight bonding model have suggested that $SrIrO_3$ should be an exotic semimetal induced by the delicate interplay between the SOI and electron correlations[27-29], in which a Dirac nodal ring near the $U$ point would render a non-trivial topological semimetallic state[27,28]. Such an exotic semimetallic state was suggested to be crucial in designing $SrIrO_3$ based artificial superlattices with novel topological properties[12,28]. However, although some optical spectroscopy and transport measurements have shown signs of semimetallic properties[2,30-32], the straightforward information on the possible semimetallic state of SrIrO$_3$, particularly its low-lying electronic structure, is still lacking to date.

Before achieving such crucial information, researchers have to overcome two major obstacles. On the one hand, the bulk SrIrO$_3$ prefers to crystallize the *6H*-hexagonal structure rather than the perovskite phase, and only the polycrystalline perovskite SrIrO$_3$ can be obtained under high pressure[33,34]. To date, there have been no bulk single crystals of perovskite SrIrO$_3$ available yet. On the other hand, while its quasi-two-dimensional analogues Sr$_2$IrO$_4$ and Sr$_3$Ir$_2$O$_7$ have been comprehensively studied using powerful electronic structure probing techniques, e.g., scanning tunneling spectroscopy (STS) and angle-resolved photoemission spectroscopy (ARPES)[1,25,35], the inability to cleave the pseudocubic structured SrIrO$_3$ prevents even a basic understanding of its low-lying electronic structure.

Aiming at the above difficulties, we applied reactive oxide molecular beam epitaxy (OMBE) to synthesize high-quality orthorhombic perovskite SrIrO$_3$ (001) films on (001) orientated SrTiO$_3$ single-crystal substrates, and then performed *in-situ* ARPES to study the low-lying electronic structure of films. Our data reveal that the hole-like and electron-like bands around *R* and *U(T)* of the orthorhombic perovskite Brillouin zone (BZ) simultaneously intersect the Fermi level ($E_F$), providing an unambiguous evidence of the semimetallic ground state in this material. These bands crossing $E_F$ are extremely narrow and there exists a pronounced mixing between the $J_{eff} = 1/2$ and $J_{eff} = 3/2$ characters therein, which is in sharp contrast to the predictions in the strong SOI limit. In addition, our photoemission results reveal that there is an evident gap between the electron- and hole-like bands near the zone boundary, indicating that the predicted Dirac degeneracy protected by the crystal-symmetry is actually lifted in the perovskite SrIrO$_3$ thin film.

## Results

**Valence band structure and Fermi-surface topology of perovskite SrIrO$_3$.**

Fig. 1(a) illustrates the three-dimensional BZ of the orthorhombic perovskite structured SrIrO$_3$, compared to that of the simple cubic lattice. Since IrO$_2$ octahedra are rotated around the *c* axis and tilted around the [110] axis, epitaxial SrIrO$_3$ has an orthorhombic perovskite structure and one orthorhombic unit cell contains four formula units with the space group of *Pbnm*. Thus, the two-dimensional projection (highlighted by the black thick lines) is reduced by half in the momentum space. More information of these epitaxial films can be found in the Supplementary Information. The left panel of Fig. 1(b) shows the second derivative of the valence band of epitaxial SrIrO$_3$ with respect to binding energy, which is in a qualitative agreement with our density functional theory (DFT) calculations (black lines appended). The right panel illustrates the angle-integrated spectrum of the valence band. By comparing with DFT calculations, we can attribute features between -7.0 eV and -2.0 eV mostly to O 2*p* states, while the Ir $t_{2g}$ orbitals mainly distribute spectral weight from -2.0 eV to 0 eV. Distinct from the fully gapped Sr$_2$IrO$_4$ and Sr$_3$Ir$_2$O$_7$[1,25], there still exists significant spectral weight in the vicinity of $E_F$ for SrIrO$_3$, consistent with the more metallic behavior of this material[2,36]. For states near $E_F$, it is worth noting that the $J_{eff}$ = 1/2 and 3/2 manifolds of SrIrO$_3$ are not well separated as in its two-dimensional counterpart Sr$_2$IrO$_4$[1], and there is a substantial mixing of $J_{eff}$ = 1/2 and $J_{eff}$ = 3/2 states.

Fig. 1(c) shows the photoemission intensity map of the epitaxial SrIrO$_3$ along with the underlying binding energy versus momenta spectra. The Fermi surface (FS) consists of a four-pointed-star shaped pocket surrounding the corner and the circular pockets around the reduced BZ boundary. This fermiology is reminiscent of the isoenergy spectral map (around 0.4 eV binding energy) of Sr$_2$IrO$_4$[1,35]. Particularly, the four-pointed-star pocket around the zone corner and the underlying hole-like band dispersion look rather like those of Sr$_2$IrO$_4$, implying the similar low-lying band characters for this series of materials despite of their different chemical potentials. While, the MIT introduces some unexpected changes of the band structure as well. For example, the electron pocket around the *X(U)* point has never been reported before even in the isoenergy spectral maps of Sr$_2$IrO$_4$ and Sr$_3$Ir$_2$O$_7$[1,25,35]. Although our photoemission data taken with He I photons (21.2 eV) can only probe one $k_z$ plane but not the complete three-dimensional BZ, the simultaneous intersection of $E_F$ by hole- and electron-like bands is one straightforward evidence that the perovskite SrIrO$_3$ has a semimetallic ground state, which has long been predicted to be caused by the delicate interplay between the strong SOI and electron correlations on the itinerant side of the RP series. This semimetallicity is also consistent with previous report on the transport measurements of epitaxial SrIrO$_3$ films[30-32]. Here, we notice the "patch-like" Fermi pocket for SrIrO$_3$, which is in contrast to the sharp FSs of Sr$_2$IrO$_4$ and Sr$_3$Ir$_2$O$_7$. This is probably induced by the intrinsic $k_z$ broadening effects arising from the short photoelectron mean free path and the sizable band dispersion along the $k_z$ direction in this pseudocubic compound.

### $k_z$ determination in the photoemission spectroscopy measurements.

Since epitaxial SrIrO$_3$ thin films have the orthorhombic perovskite structure, the $k_z$ dispersion of these films is expected to be substantial. Therefore, it is crucial in our work to verify that the $k_z$ plane we probed is rather close to $Z$-$U$-$R$-$T$ plane, so that we can compare our data directly with theoretical predictions.

Using the free-electron final-state model[37], in which the inner potential was set to 11 eV [the typical inner potential value for perovskite transition metal oxides[38-40], and this value have been used in a recently published photoemission work on epitaxial SrIrO$_3$ films[41]], we can estimate the $k_z$ values probed with He I (21.2 eV) and He II (40.8 eV) photons and then append both of them on the $k_z$–$k_x$ plane [the light blue one highlighted in Fig. 2(a)]. As illustrated in Fig. 2(b), photons with these two energies accidently probe the equivalent $k_z$ planes (the $Z$-$U$-$R$-$T$ high-symmetry plane) in the Brillouin zone.

Actually, the predicted Dirac node discussed below is expected to be located in a rather narrow $k_z$ window, which requires that our ARPES cut position should be accurately close to the $Z$-$U$-$R$-$T$ plane to probe this feature. The ARPES measurement would not even probe such Dirac cone like bands once the cut positon of $k_z$ is out of this range. However, later it will be shown that we have successfully detected such Dirac line nodes, which are in a good agreement with our and other theoretical predictions. This finding strongly supports our judgement that we have indeed probed the electronic structure around $U$. In addition, all the measured band dispersions with

He II rather resemble those obtained with He I photons except for some photoemission intensity changes [Figs. 2 (c-d)]. This finding indicates that both He II and He I photons should probe the equivalent $k_z$ planes in the BZ. Even if slightly deviating from this inner potential of 11 eV, the ARPES cuts taken with He I & II photons would not be possible to be located in the equivalent planes according to the free-electron final-state model. In that case, we should have observed the evident band dispersing through the comparison of ARPES data taken with He I & II photons, respectively, considering the pronounced three-dimensional nature of the band structure of $SrIrO_3$. Clearly, this is inconsistent with our experimental results. This finding again confirms that both ARPES measurements with He I & II photons indeed probe the *Z-U-R-T* high-symmetry plane and our choice of 11 eV inner potential are reliable.

**Details of band dispersions along the high-symmetry directions.**

To investigate more details of the semimetallic ground state of epitaxial $SrIrO_3$ films, we further examined the band dispersions along several high-symmetry directions of the unfolded BZ, as indicated in the right panel of Fig. 3(j).

Fig. 3(a) show the photoemission data taken along *R-Z* direction (Cut #1). Around the *Z* point, two hole-like bands can be resolved through the second derivative with respect to energy [Fig. 3(d)], and we can assign them as bands *α* and *β*, respectively. As further shown by the corresponding momentum distribution curves (MDCs) [Fig. 3(g)], *α* just barely intersects $E_F$, forming the small hole pocket around the *Z* point as aforementioned in Fig. 1(c). While, the *β* band sinks slightly below $E_F$, with the top centering at the binding energy of ~20 meV. Here, both bands are degenerate at the *Z* point, which is protected by the lattice symmetry. Along this direction, both bands *α* and *β* further disperse backwards to $E_F$ around the zone corner and then cross $E_F$ [Fig. 3(d), and 3(g)]. We note that this is in a qualitative agreement with the previous reports on $Sr_2IrO_4$ and $Sr_3Ir_2O_7$, in which the tops of both valence bands appear at the zone corner rather than the center. Surrounding the *R* point, one more hole-like band with larger Fermi crossing (*γ*, marked by the orange dashed line) can be distinguished, which develops the outer four-pointed-star like Fermi pocket around the zone corner. Besides, the second derivative plot [Fig. 3(d)] clearly shows that there exist two weak hole-like features (*ω* and *τ*) with band tops at the binding energy of ~200 meV, which surprisingly replicate all features of bands *β* and *γ*, respectively. They will be discussed more in another work (Z. T. L., D. W. S. *et al.,* manuscript in preparation).

For the *U-T* direction (Cut #2), a parabolic electron-like feature (assigned as *δ*) crosses $E_F$, forming the circular electron pocket surrounding *U(T)*, as illustrated by both the photoemission intensity [Fig. 3(b)] and second derivative plots [Fig. 3(e)]. The corresponding MDCs [Fig. 3(h)] further demonstrate that its band bottom is located at around 40 meV below $E_F$. Here, the band top of *γ* is tight against the band bottom of *δ*, which rather resembles the predicted Dirac-like dispersion around *U(T)*[27,28]. However, we note this Dirac cone is obviously gapped from the node point. And here, the spectra look rather symmetrical along *U-T* direction though the *U* and *T* points are not equivalent for a single crystalline $SrIrO_3$. As for epitaxial $SrIrO_3$ films, the square $SrTiO_3$ (100) (*a* = *b* ~ 3.905Å) substrates underneath could well account for this symmetry. For simplicity, we will unify the *U(T)* points as *U* thereafter.

Figs. 3 (c), (f) and (i) illustrate the data along another high symmetry direction *R-A* (Cut #3), which are reminiscent of those along *Z-R* except for some differences in the relative spectral intensity caused by matrix elements. We note that the epitaxial $SrIrO_3$ films are of the orthorhombic perovskite structure, in which the band structure would be folded along the diagonal lines of the tetragonal BZ [Fig. 1(a)]. Consequently, the *R-A* direction should be equivalent to *R-Z* in the two-dimensional projection of the orthorhombic BZ. Such band folding phenomena have been reported in $Sr_2IrO_4$ and $Ba_2IrO_4$ single crystals or films as well[1,42].

**Comparison between the DFT band structure and experimental results.**

In order to better understand the electronic structure of $SrIrO_3$, we have performed the LDA+SO+U calculation with different U values to compare with our ARPES result, as shown in Figs. 4(a-c). We find that the only effect of applying different U for $SrIrO_3$ is to induce some band energy shifts, in sharp contrast to the case of the insulating $Sr_2IrO_4$, in which an evident gap between the $J_{eff}$ = 1/2 and 3/2 would be open with larger U value. The calculation with U = 0 eV [Fig. 4(a)] rather than those with U = 1 eV and 2 eV [Figs. 4(b-c)] is more in line with the experimental band structure, particularly around the Dirac cone at the *U* point. This result indicates that electron correlations in perovskite SrIrO$_3$ is relatively weak, consistent with the factor that the screening would be larger in three-dimensional systems than that in two-dimensional Sr$_2$IrO$_4$. We find that our calculation is rather similar to previous theoretical works[2,5,27-29]. Particularly, all these calculations predict the Dirac cone like bands around the *U* point. However, as shown in Fig. 4(a), our calculation explicitly predicts that the degeneracy of this Dirac node is lifted, in better agreement with our experiments [Fig. 4(d)].

To analytically deduce the renormalization factor, we focus on *β* band since its large portion can be observed in the experiment. By applying parabolic fitting to the band dispersion of *β* around the zone corner, its band top is estimated to be around 20 meV above $E_F$, and the total bandwidth is approximately 290 meV. Compared with calculations (360 meV), the renormalization factor is determined to be 1.24. The small value is consistent with the relatively weak electron correlations in these 5*d*-electron systems.

Qualitatively, most characteristic dispersions of bands *α*, *β* and *δ* near $E_F$ can be verified by the DFT calculation. Particularly, the predicted Dirac cone like feature around *U* can be well resolved. However, there still exist some inconsistencies between them. For example, along the *Z-U* direction, *α* and *β* bands from ARPES results do not intersect, while the DFT only predicts two degenerate bands. Besides, we find several bands which have been probed by experiments but cannot be well captured in the calculation, including part of *γ* and the replica bands *ω* and *τ*. We note that *γ* crosses with *α* and *β* but does not show any hybridization effects, implying it might be a shadow band that is a replica of some band transferred by specific momentum vector.

**Gapped bands around the *U* point.**

One of the hottest topics on iridates is that the perovskite SrIrO$_3$ based artificial superlattices might be the long-sought-after oxide topological insulators with relatively large band gap[5-7,12,28]. This non-trivial topological property was thought to be inseparably linked with the predicted nodal ring around the $U$ point of SrIrO$_3$. As shown in the Fig. 3(b), the band dispersion around $U$ rather resembles the predicted linear Dirac cone crossing. However, the corresponding second derivative plot in Fig. 3(e) demonstrates that the electron- and hole-like features do not touch each other but are isolated by a gap of around 30 meV, and their dispersions deviate from the characteristic linearity of a typical Dirac cone. This means that the Dirac degeneracy protected by the mirror-symmetry in the theory might have been lifted in epitaxial perovskite SrIrO$_3$ thin films.

To further validate this finding, we carried out a comprehensive study of the low-lying band structure near $U$. Typical second derivative plots and the corresponding EDCs spectra taken along some other high-symmetry directions are summarized in Fig. 5. Similar to the data taken along $U$-$T$ direction as shown in Figs. 3[(b), (e), (h)], there is a more accurate estimated gap of about 28 meV determined by the corresponding EDCs separated between the upper and lower Dirac cones around $U$ along both cut #1 ($U$-$Z$) and cut #2 ($U$-$R$) [Figs. 5 (a-d)]. This is as well confirmed by our zoom-in DFT calculations, as illustrated in Figs. 5 (e-f). Overall, we could conclude that the low energy band structure of epitaxial perovskite SrIrO$_3$ films is not so consistent with the prediction based on the tight bonding model[28], and the Dirac line nodes are gaped though the mirror-symmetry of the lattice is still conserved.

## Discussion

One intriguing finding of our work is that the bands crossing $E_F$ in epitaxial SrIrO$_3$ disperse only 290 meV across nearly the whole BZ, indicating a rather narrow bandwidth of the $J_{eff} = 1/2$ bands. This bandwidth is close to that of the lower Hubbard band of Sr$_3$Ir$_2$O$_7$ (150~200 meV)[25], but even narrower than that of Sr$_2$IrO$_4$ (~500 meV)[1]. Our findings show that the bandwidth of the $J_{eff} = 1/2$ band does not increase evidently for SrIrO$_3$ as expected[1,2]. Besides, for the bands in the vicinity of $E_F$, neither our photoemission data nor DFT calculations show the clear energy gap between the $J_{eff} =$

1/2 and $J_{eff}$ = 3/2 states, and there exists a significant mixing of these two states. This result is qualitatively consistent with the cases of $Sr_3Ir_2O_7$[25] and $Sr_3CuIrO_6$[43], but in contrast to the expectations of the strong SOI limit.

Such narrow bandwidth of the $J_{eff}$ = 1/2 band and the significant mixing of $J_{eff}$ = 1/2 and 3/2 states (Fig. 3) pose a strong challenge to the current prevalent explanation for the MIT occurring in the RP series $Sr_{n+1}Ir_nO_{3n+1}$. In this model, the cooperative interplay between the moderate electron correlations and strong SOI would result in the Mott MIT in the narrow half-filled $J_{eff}$ = 1/2 band of $Sr_2IrO_4$. With increasing neighboring Ir atoms in the $Sr_{n+1}Ir_nO_{3n+1}$ series, the bandwidth $W$ of the $J_{eff}$ = 1/2 band would keep rising, and the metallicity would restore until the $W$ is comparable to or even larger than the electron correlation $U$, which is the case of $SrIrO_3$[2]. This departure from the idealized $J_{eff}$ = 1/2 and 3/2 picture and the MIT in the RP series $Sr_{n+1}Ir_nO_{3n+1}$ might be induced by the extremely large crystal field splitting comparable to the SOI in this system[41].

Moreover, according to the previous tight binding calculations, Jean-Michel Carter *et al*. [Ref. 28] conjectured that the line node near *U* point should not be gapped unless the mirror symmetry of the lattice is broken. They suggested that the breaking of this symmetry by introducing a staggered potential or spin-orbit coupling strength between alternating layers would turn this line node into a pair of three-dimensional nodal points, and consequently induce a transition to a strong topological insulator[28]. In our experiments, the $SrTiO_3$ (100) substrate itself and the induced compressive strain would not expected to break such a *Pbnm* mirror-symmetry of the $SrIrO_3$ lattice. However, neither the photoemission data nor calculations are consistent with that proposal. Recently, some works proposed that the Dirac line node in epitaxial $SrIrO_3$ film is actually protected by the *n*-glide symmetry operation, which could be selectively broken by different hetero-epitaxial structures. The Dirac line node degeneracy could be lifted without breaking the *Pbnm* mirror-symmetry[5,6]. In this work, the authors applied (110)-oriented $GdScO_3$ substrates to fabricate epitaxial $SrIrO_3$ films, and they calculated that the *0.6%* tensile strain would lead to a small gap of 5 meV. Instead, our films were grown on (100)-oriented $SrTiO_3$, which leads to a *1.54%* compressive strain to the pseudo-cubic SrIrO$_3$. Based on the parameters of our own heterostructure, the gap size is calculated to be around 20 meV, remarkably close to our experimental result 28 meV. In this regard, our SrIrO$_3$ (100)/SrTiO$_3$ (100) heterostructure would have inevitably broken the *n*-glide symmetry, which would consequently result in the lifting of Dirac nodes in the perovskite SrIrO$_3$ thin films.

Though our DFT calculation well reproduces the gapped Dirac nodes around the *U* point in epitaxial SrIrO$_3$ film. However, our experimental findings are still not in complete agreement with the theory, and there still exist some inconsistencies between our band structure calculation and experiments. On the one hand, the delicate interplay between the comparable SOI and electron correlations would perplex the precise prediction of the band structure calculation. On the other hand, in the real calculation, it is difficult to fully consider all the subtle lattice distortions in epitaxial films, which would affect significantly the final band structure calculations[5,40]. Thus, further more sophisticated theoretical considerations are desired.

In summary, we have applied the combo of OMBE and *in-situ* ARPES to systematically study the low-lying electronic structure of the orthorhombic perovskite SrIrO$_3$ films. Our data show that the hole-like bands around *R* and the electron-like bands around *U* intersect $E_F$ simultaneously, providing the direct evidence of the semimetallic ground state in the perovskite structured SrIrO$_3$. In addition, we find that the bandwidth of the states in the vicinity of $E_F$ is extremely small, and there exists a pronounced mixing between the $J_{eff} = 1/2$ and $J_{eff} = 3/2$ bands, which is in sharp contrast to the predictions in the strong SOI limit. Both our photoemission results and DFT calculations reveal that there is an evident gap between the electron-and hole-like cones near the zone boundary, which lifts the predicted Dirac line node degeneracy of perovskite SrIrO$_3$. Our findings pose strong constraints on the current theoretical models for the 5*d* iridates, and thus call for a further refinement.

**Note Added**

At this work was made public through posting as an e-print (arXiv: 1501.00654), we note that, a photoemission work, which reports some similar results as part of our findings and was carried out independently by another group, was published: see Ref. [41].

## Methods:

### Sample synthesis

Perovskite structured $SrIrO_3$ films of 25 unit cells (~10 nm, referenced to the pseudocubic cell) were deposited on (001) $SrTiO_3$ substrates using a DCA R450 OMBE system at a substrate temperature of 560 °C in a background pressure of $2\times10^{-6}$ torr of distilled ozone. The growth temperature was verified by optical pyrometry. The lattice constant of $SrTiO_3$ substrate is 3.90 Å, which leads to a 1.54% compressive strain to the pseudocubic $SrIrO_3$ (3.96 Å). Strontium and iridium were evaporated from an effusion cell and an electron beam evaporator, respectively, and the co-deposition method with both elements' shutters being open for the duration of the growth was applied. The strontium and iridium fluxes were approximately $1.2\times10^{13}$ atoms/(cm$^2$s), which were both checked before and after the deposition by a quartz crystal microbalance. During the growth, the in-situ reflection high-energy electron diffraction (RHEED) intensity oscillations and patterns were collected to monitor the overall growth rate and surface structure, respectively. The films' structure was finally ex-situ examined by X-ray diffraction (XRD) using a high-resolution Bruke D8 discover diffractometer.

### ARPES measurement

Thin films were transferred to the combined ARPES chamber for measurements immediately after the growth through an ultrahigh vacuum buffer chamber (~$1.0\times10^{-10}$ torr). This ARPES system is equipped with a VG-Scienta R8000 electron analyzer and a SPECS UVLS helium discharging lamp. The data were collected at 15 K under ultrahigh vacuum of $8.0\times10^{-11}$ torr. The angular resolution was 0.3°, and the overall energy resolution was set to 15 meV. During the measurements, the films were stable and did not show any sign of degradation.

**Acknowledgements**
The authors are grateful to Professor Donglai Feng for helpful discussions. This work was supported by the National Key R&D Program of the MOST of China (Grant No. 2016YFA0300204), the National Basic Research Program of China (973 Program) under Grants No. 2012CB927400, and the National Science Foundation of China under Grants Nos. 11274332, 11574337, 11525417 and 11227902. D. W. S. is also supported by the "Strategic Priority Research Program (B)" of the Chinese Academy of Sciences (Grant No. XDB04040300) and the "Youth Innovation Promotion Association CAS". Q. F. Li was supported by China Postdoctoral Science Foundation (2014M551544).

**Author contributions statement**
Z.T.L. and M.Y.L. prepared the samples. D.W.S. conceived and planned the experiments. Z.T.L., H. F.Y., Q.Y. and C.C.F. carried out the experiments. D.W.S., Z.T.L. and J.S.L. carried out data analysis. D.W.S., Z.T.L. and Z.W. discussed the results. X.G.W., Q. F. L. and W.L. conducted band calculations. D.W.S., Z.T.L. and X.G.W. wrote the paper. All authors reviewed the manuscript.

**Additional information**
Competing financial interests: The authors declare no competing financial interests.

**Supplementary information**

**Correspondence and requests** for materials should be addressed to **D. W. S.** (email: dwshen@mail.sim.ac.cn).

# Figure

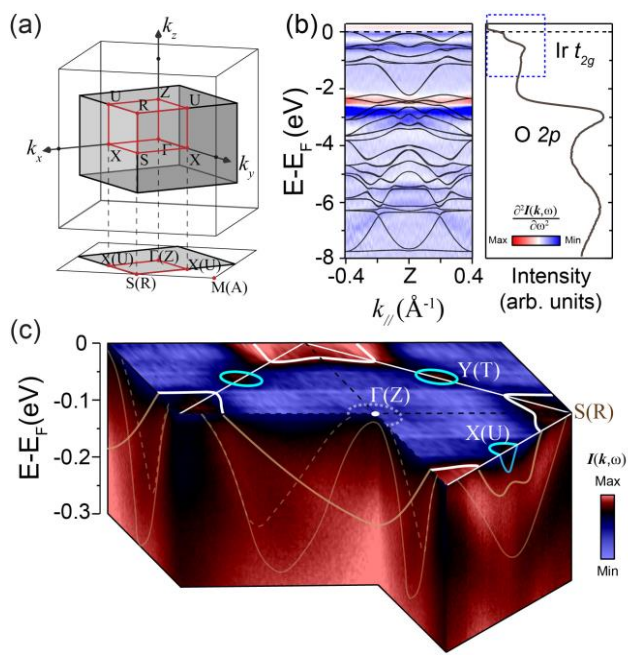

Figure 1| **Valence band and Fermi-surface topology of SrIrO$_3$ films.** (a) The three-dimensional Brillouin zone of orthorhombic perovskite SrIrO$_3$ and its projection in the $k_x$ and $k_y$ plane. (b) The comparison between the second derivative plot of the valence band and DFT calculation of SrIrO$_3$ (left panel), and the corresponding angle-integrated spectrum (right panel). (c) The photoemission intensity map integrated over [$E_F$ - 10 meV, $E_F$ + 10 meV] of the epitaxial SrIrO$_3$ along with the underlying binding energy versus momenta spectra. The data were taken with He I (21.2 eV) photons.

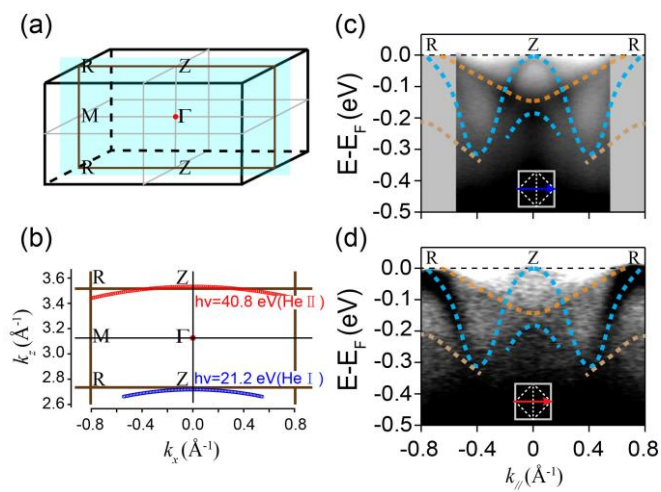

Figure 2| **Schematics illustrating the sites of scan projected to $k_z$-$k_x$ plane in three-dimensional Brillouin zone.** (a) The Brillouin zone of orthorhombic perovskite

structured epitaxial SrIrO$_3$ films with compressive misfit strain. The symbols for high-symmetry momentum points ($\Gamma$, $M$, $Z$ and $R$) follow the Brillouin zone of an orthorhombic perovskite structure. (b) Sites of two photon energies (HeⅠ & HeⅡ) scan projected to $\Gamma$-$M$-$R$-$Z$ plane. (c)-(d) The photoemission intensity plots taken with He I (21.2 eV) photons and HeⅡ (40.8 eV) along $Z$-$R$, respectively.

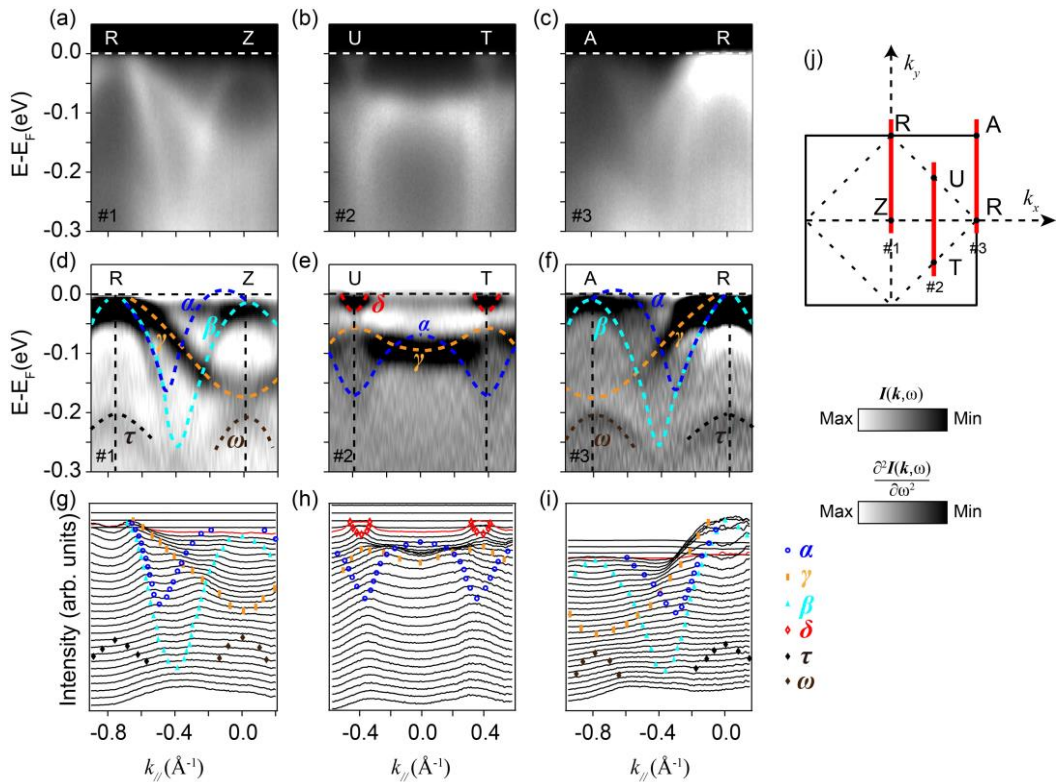

Figure 3| **Low energy electronic structure along high-symmetry directions.** (a)-(c) The photoemission intensity plots taken along $Z$-$R$, $T$-$U$, and $R$-$A$ high-symmetry directions, respectively. (d)-(f) The corresponding second derivative images with respect to the energy for these cuts. (g)-(i) The MDCs for the data in (a)-(c). All these data were taken with 21.2 eV photons, corresponding to the $Z$-$U$-$R$-$T$ $k_z$ plane. (j) The indication of the cut direction in the projected two-dimensional Brillouin zone.

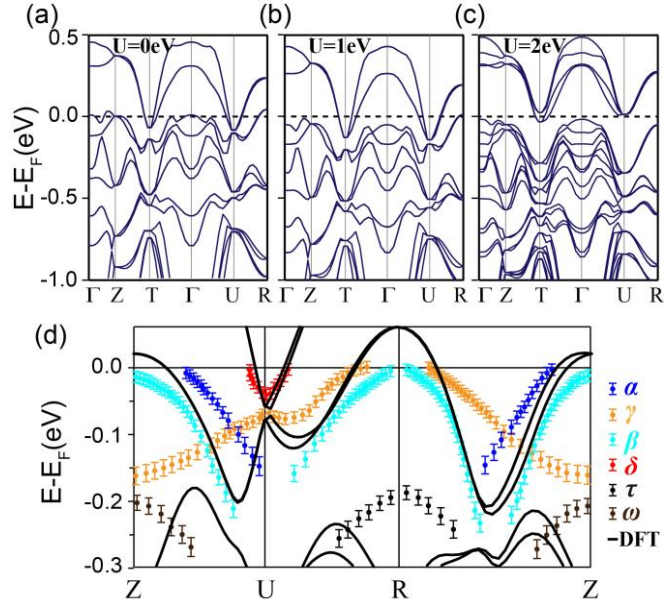

Figure 4| **Bands of DFT calculations and experimental results.** (a)-(c) Representative DFT band structures of orthorhombic perovskite SrIrO$_3$ films for (a) U = 0 eV, (b) U = 1.0 eV, and (c) U = 2.0 eV. (d) Comparison of the ARPES results and the DFT calculations (U = 0 eV) along the high-symmetry directions in epitaxial SrIrO$_3$ films. Experimental band dispersions are determined either by fitting both EDCs and MDCs of the raw data and or from peak positions in second derivative data. DFT calculations are marked by the solid lines.

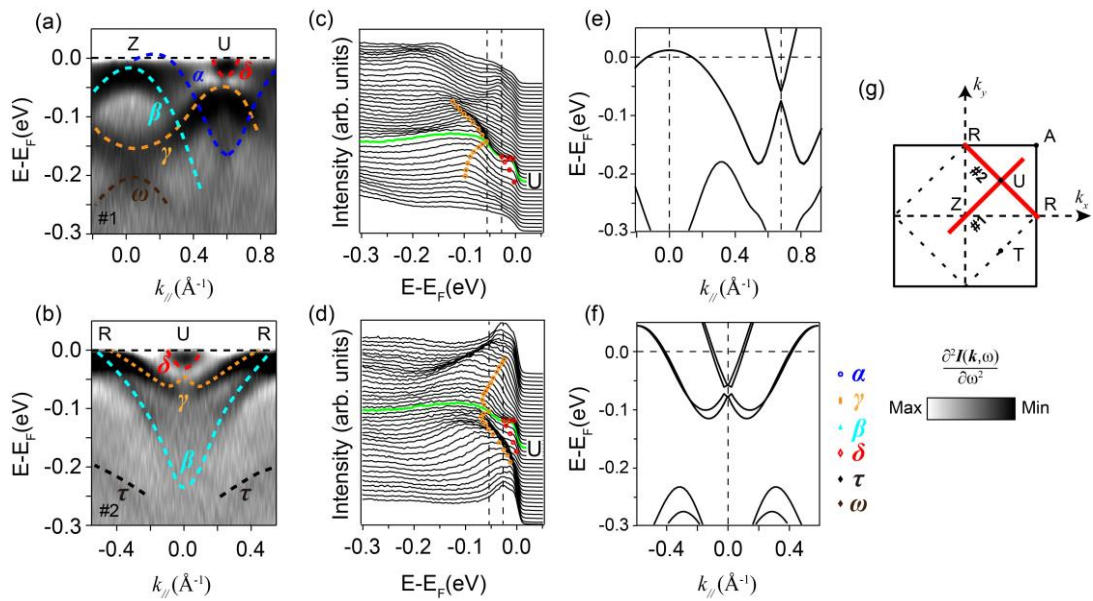

Figure 5| **Dirac line node degeneracy lifting around the *U* point.** (a), (b) The second

derivative images with respect to the energy for the photoemission data along *Z-U* and *U-R* high-symmetry directions, respectively. (c), (d) The corresponding EDCs for the photoemission data in (a) and (b), respectively. (e), (f) The calculated band dispersions along *Z-U* (cut #1) and *U-R* (cut #2) high-symmetry directions, respectively. All these data were taken with 21.2 eV photons. (g) The indication of the cut direction around U point in the projected two-dimensional Brillouin zone.